\documentclass[12pt]{article}
\usepackage{amsmath,amssymb}
\usepackage{graphicx}
\usepackage{geometry}
\def\TODAY{10 February 2011; 7 April 2011}
\title{\bf Elementary analysis of the special relativistic combination of velocities, \\
Wigner rotation, \\
and Thomas precession}
\author{{Kane O'Donnell\thanks{Email: kco61@uclive.ac.nz}
\thanks{Current address: Department of Physics \& Astronomy, University of Canterbury, New Zealand}\; 
and Matt Visser\thanks{Email: matt.visser@msor.vuw.ac.nz}}
\\ 
\\
School of Mathematics, Statistics, and Operations Research
\\
Victoria University of Wellington, New Zealand}
\date{\TODAY, \LaTeX-ed \today}

\begin{document}
\maketitle
\begin{abstract}
The purpose of this paper is to provide an elementary introduction to the qualitative and quantitative results of velocity combination in special relativity, including the Wigner rotation and Thomas precession. We utilize only the most familiar tools of special relativity, in arguments presented at three differing levels: (1) utterly elementary, which will suit a first course in relativity; (2) intermediate, to suit a second course; and (3) advanced, to suit higher level students. We then give a summary of useful results, and suggest further reading in this often obscure field.

\bigskip
\noindent
European Journal of Physics {\bf 31} (2011)  1033--1047\\
doi: 10.1088/0143-0807/32/4/016



\end{abstract}
\clearpage
\tableofcontents
\clearpage

\section{Introduction}

The problem of how to consider velocities in a special relativistic setting is fundamental to many areas of both theoretical and applied physics~\cite{moller1952, jackson1998, stapp1956, fisher1972, ferraro1999, malykin2006, ritus2007}. However, students are rarely introduced to anything beyond the most basic of results (such as the relativistic composition of parallel velocities), on account of the perceived complexity and confusion surrounding the combination of velocities in special relativity. The aim of this paper is to remove some of this confusion, and clarify the qualitative concepts associated with the relativistic combination of velocities, which we do in Section~\ref{qualitativeintro}. This includes a description of what such velocities actually ``mean'', what the Wigner rotation represents, and how this leads to the Thomas precession.

In Section~\ref{elementary} we provide derivations of certain key quantitative results, using only elementary concepts of special relativity. We begin with the simple cases of relativistically combining parallel and perpendicular velocities in Section~\ref{parperpsec}. {This section is particularly relevant for those new to such concepts. The formulae are simple and elementary to derive, yet still illustrate the fundamental issues of combining relativistic velocities, including the Wigner rotation and Thomas precession. In Section~\ref{generalsec} we use the results already obtained to consider the general combination of velocities --- that is, where the velocities are neither necessarily parallel nor perpendicular. We envisage this section to be suitable for students undertaking a first course in relativity, though proving the results of parts of Section~\ref{generalsec} involves extensive vector manipulation.

In Section~\ref{intermediate} we consider the relativistic combination of velocities using the boost matrix formulation of special relativity. Whilst in principle this is no more complex than our elementary derivations of Section~\ref{elementary}, a familiarity with the boost matrix representation is assumed, and hence this section will likely be suitable for students undertaking a second course in relativity.

In Section~\ref{advanced} we briefly outline how the spinor formulation of special relativity can reproduce the results we have already obtained. This section is only suitable for those already familiar with the spinorial representation of Lorentz transformations, and hence is likely to be accessible mainly for more advanced students.

Lastly, in Section~\ref{summary} we give a summary of important (and often equivalent) formulae in this field, and in Section~\ref{furtherreading} we provide references for further reading. 

\section{Qualitative introduction}\label{qualitativeintro}
\subsection{Relativistic combination of velocities}
To begin with, consider Alice and Bob, each traveling in a spaceship somewhere in the vicinity of Earth, as illustrated in Figure~\ref{figureintro}. Unfortunately, due to equipment malfunction, mission control cannot directly observe the velocity of Bob. Nonetheless, Alice is able to measure the velocity of Bob to be $\vec{v}_2$, and mission control can measure the velocity of Alice to be $\vec{v}_1$. The key question surrounding the relativistic combination of velocities is how we deduce the velocity $\vec{v}_{21}$ of Bob, as seen by mission control, using the velocities $\vec{v}_1$ and $\vec{v}_2$. (Note that from the beginning, we must be clear that $\vec{v}_2$ is measured in Alice's rest frame whilst $\vec{v}_1$ and $\vec{v}_{21}$ are measured in mission control's rest frame.) As shown in Section~\ref{generalsec} we may indeed derive a simple formula for this velocity $\vec{v}_{21}$, and it is this quantitative result that embodies what we mean by the relativistic combination of velocities $\vec{v}_1 $ and $\vec{v}_2$.

\begin{figure}[!htb]
\begin{center}
\includegraphics[width=0.4\textwidth]{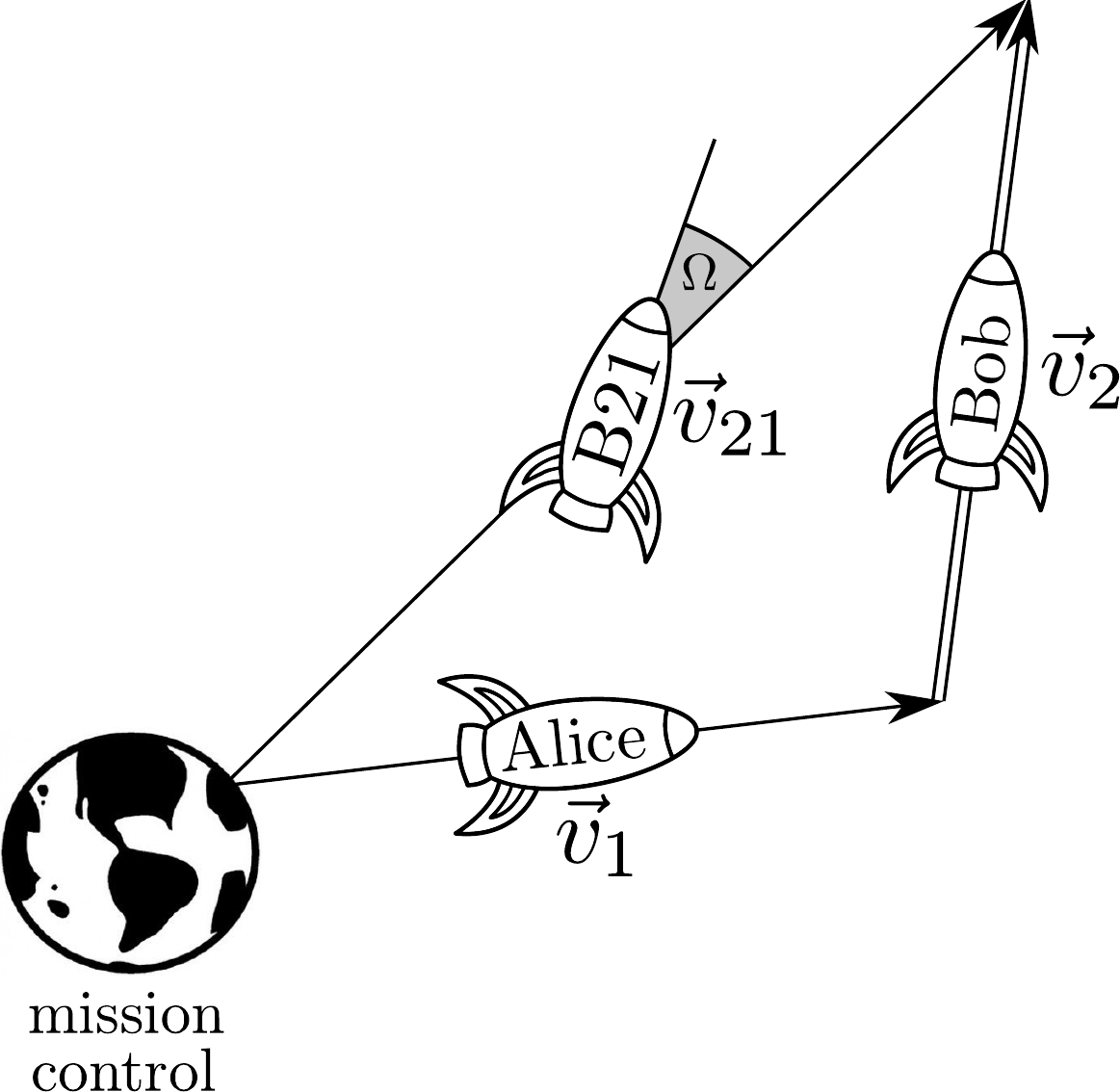}
\end{center}
\caption{\textit{The common (and misleading) depiction of the combination of velocities. Mission control sees Alice moving with velocity $\vec{v}_1$, and Alice sees Bob moving with velocity $\vec{v}_2$ (shown as a double line to indicate this is in Alice's frame). Mission control observes Bob as the spacecraft labeled B21, and to be moving at velocity $\vec{v}_{21}$, but pointing in a direction rotated by the Wigner rotation angle $\Omega$. From mission control's perspective, Bob appears to be ``sliding'' sideways in the direction $\vec{v}_{21}$.}}
\label{figureintro}
\end{figure}

Note that Figure~\ref{figureintro} is somewhat misleading in that it treats the velocity vectors like Euclidean displacement vectors --- in reality, they need not be linked ``head-to-tail''. Nonetheless this presentation makes it clear which velocities are being combined, and hence remains qualitatively useful.

\subsubsection{Wigner rotation}
The relativistically combined velocity $\vec{v}_{21}$ cannot be interpreted as directly as we are used to. Whilst mission control observes Bob to be traveling with velocity $\vec{v}_{21}$, they will observe him to be pointing at an angle $\Omega$ to $\vec{v}_{21}$, as illustrated in Figure~\ref{figureintro}. That is, Bob's frame of reference appears to mission control to be rotated by an angle $\Omega$, which is known as the Wigner rotation angle.

\begin{figure}[!htb]
\begin{center}
\includegraphics[width=0.5\textwidth]{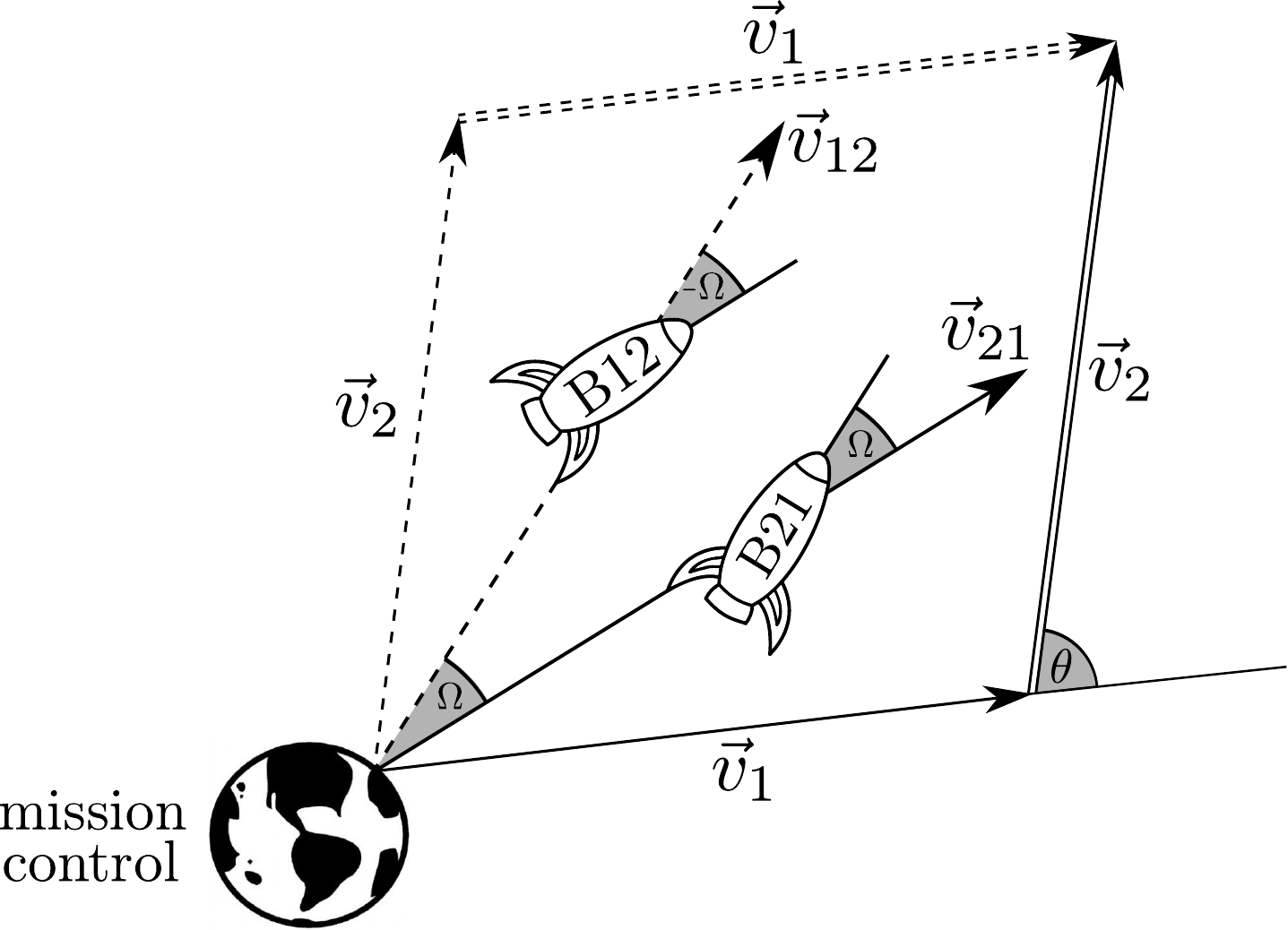}
\end{center}
\caption{\textit{A more correct interpretation of the relativistic combination of velocities. The solid lines indicate the case where Alice has velocity $\vec{v}_1$ as measured by mission control, and Bob has velocity $\vec{v}_2$ as measured by Alice, resulting in mission control seeing Bob moving with velocity $\vec{v}_{21}$ (that is, in the spacecraft B21). The dashed lines indicate the naively ``symmetrical'' case, where Alice has velocity $\vec{v}_2$ as measured by mission control, and Bob has velocity $\vec{v}_1$ as measured by Alice, resulting in mission control seeing Bob moving with velocity $\vec{v}_{12}$ (that is, in the spacecraft B12). In addition, the Wigner rotation angle, $\pm{}\Omega$ in each case, is also the angle between $\vec{v}_{12}$ and $\vec{v}_{21}$. The angle $\theta$ is that between $-\vec{v}_1$ and $\vec{v}_2$, as measured by Alice.}}
\label{figurev12v21}
\end{figure}

One may also consider the apparently ``symmetrical'' case of the combined velocity $\vec{v}_{12}$, where we imagine instead that Bob has a velocity $\vec{v}_1$ as seen by Alice, and Alice has a velocity $\vec{v}_2$ as seen by mission control, as illustrated in Figure~\ref{figurev12v21}. In standard Galilean relativity, we would predict quite rightly that $\vec{v}_{12}=\vec{v}_{21}$. However this is \textit{not} the case when considering the relativistic combination of velocities, as although $||\vec{v}_{12}||=||\vec{v}_{21}||$, they do not point in the same direction --- there is some angle $\Omega$ between them. As we will show, this is also the aforementioned Wigner rotation angle. Indeed, the observation that the Wigner rotation angle corresponds to the angle between $\vec{v}_{12}$ and $\vec{v}_{21}$ can be further extended: In the case of mission control seeing Alice moving with velocity $\vec{v}_1$, and Alice seeing Bob moving with velocity $\vec{v}_2$, then mission control will see Bob traveling at velocity $\vec{v}_{21}$, but pointing in the direction of $\vec{v}_{12}$, as is illustrated in Figure~\ref{figurev12v21}. A similar argument applies for $\vec{v}_{12}$.

\subsubsection{Thomas precession}
\def\rmd{\mathrm{{d}}}

The Thomas precession is a consequence of the Wigner rotation, and arises when one considers the case of Bob experiencing some form of centripetal acceleration. To set up a suitable scenario, let us assume that Alice and Bob are traveling off together to explore the moon, and hence, at some time $t$, are both traveling at velocity $\vec{v}_1$ as seen by mission control. Hence Alice observes Bob at rest in her frame. This is illustrated in Figure~\ref{figurethomassetup}a). 

However Bob's boosters suddenly fail, and hence he falls into a circular orbit around Earth, while Alice continues to travel in a straight line toward the moon.\footnote{Alice is using her boosters so as not to fall into orbit, and hence is not accelerating with respect to mission control. We should also be clear that we are considering the case of Newtonian gravity, and only for the pedagogical purpose of providing a centripetal acceleration in this example.  In fact any force applied through the centre of mass would do the job, such as (for example) a string attached to a gimbal at the centre of mass. For an electron one might like to consider an electromagnetic force.}
Hence Alice now measures Bob to have some velocity $\rmd\vec{v}_2$ at a later time $\rmd{}t$. This is analogous to the situation depicted in Figure~\ref{figureintro}, except now $\vec{v}_2$ becomes the infinitesimal $\rmd{}\vec{v}_2$. At the time $t+\rmd{}t$, we think of Bob's velocity relative to mission control as $\vec{v}_{21}=\rmd{}\vec{v}_2\oplus\vec{v}_1$, and his frame to be rotated by the infinitesimal Wigner rotation angle $\rmd{}\Omega$, as illustrated in Figure~\ref{figurethomassetup}b). The associated rate of change of the Wigner rotation angle $\rmd{}\Omega/\rmd{}t$ (that is, how fast Bob's frame is rotating relative to mission control's) is called the Thomas precession rate. The actual Thomas precession $\Omega_T$ is the total Wigner rotation turned through if Bob carries out a complete orbit. That is,
\begin{equation}
 \Omega_T=\oint_C{\frac{\rmd{}\Omega}{\rmd{}t}}\rmd{}t
\end{equation}
for any closed curve $C$ in velocity-space.

\begin{figure}[!htb]
\begin{center}
\includegraphics[width=0.8\textwidth]{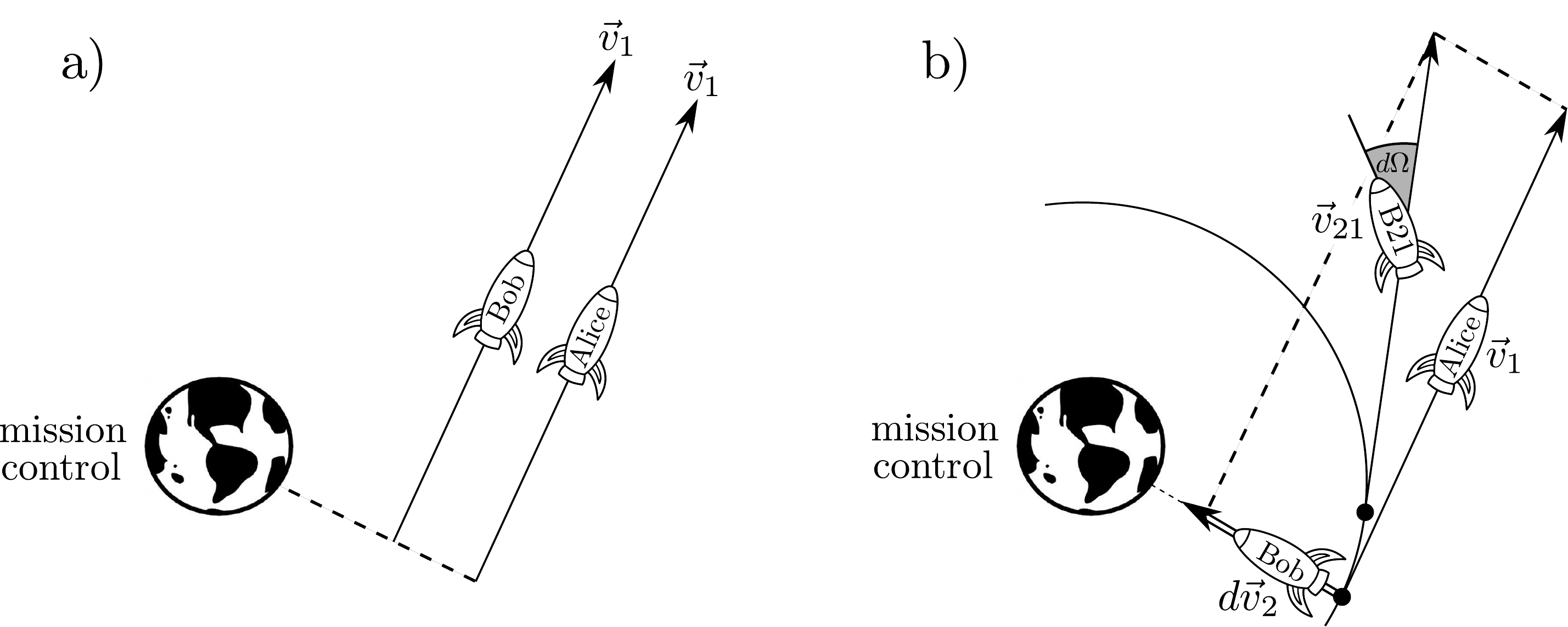}
\end{center}
\caption{\textit{At time $t$ mission control sees both Alice and Bob traveling at velocity $\vec{v}_1$, as shown in subfigure a). However Bob's boosters fail, and he falls into Earth's orbit, so at a time $\rmd{}t$ later, Alice measures Bob to have a velocity $\rmd{}\vec{v}_2$. Mission control now sees Bob (labeled as spacecraft $B21$) to be moving with velocity $\vec{v}_{21}=\rmd{}\vec{v}_2\oplus\vec{v}_1$, and his direction rotated by the infinitesimal Wigner rotation angle $\rmd{}\Omega$, as shown in subfigure b).}}
\label{figurethomassetup}
\end{figure}

\section{Elementary level discussion}\label{elementary}
\subsection{Parallel  velocities}\label{parsec}
To begin the discussion, we consider the relativistic combination of velocities for the special cases of parallel and perpendicular velocities $\vec{v}_1$ and $\vec{v}_2$, as illustrated in Fig~\ref{figureparperp}a) and Fig~\ref{figureparperp}b) respectively. However, as the relativistic combination of parallel velocities formula is usually given in textbooks (see for example \cite{moller1952} or \cite{jackson1998}), we merely state the well-known result:
\begin{equation}
 \vec{v}_{21}=\vec{v}_{12}=\frac{\vec{v}_1+\vec{v}_2}{1+\vec{v}_1\cdot\vec{v}_2}\;,
 \label{parallel}
\end{equation}\\
where we have set $c=1$. It is important to note that in this case the direction and magnitude of the two combined velocities $\vec{v}_{21}$ and $\vec{v}_{12}$ are the same, and hence there will be no resulting Wigner rotation or Thomas precession. It also illustrates that Thomas precession can indeed only occur for centripetal motion, where $\vec{v}_1$ and $d\vec{v}_2$ are not collinear.

\begin{figure}[!htb]
\begin{center}
\includegraphics[width=0.5\textwidth]{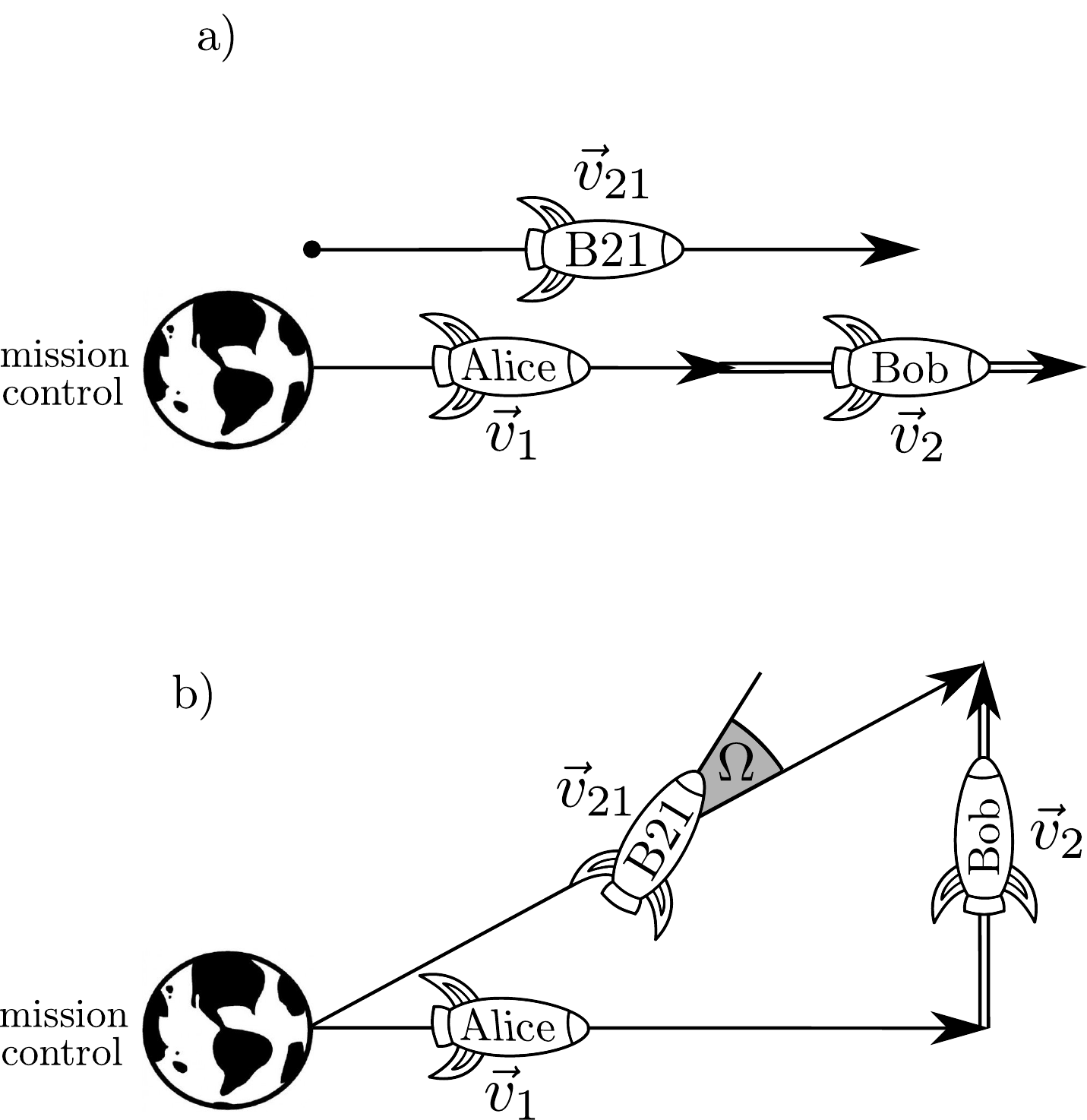}
\end{center}
\caption{\textit{Parallel and perpendicular relativistic combination of velocities is illustrated in subfigures a) and b) respectively.}}
\label{figureparperp}
\end{figure}

\subsection{Perpendicular velocities}\label{parperpsec}

The formula for the relativistic combination of perpendicular velocities can be derived in a similar manner as for the parallel case. Here we will use the elementary concepts of time dilation and length contraction. A more explicit Lorentz transformation calculation can easily verify the following results.

Consider the case illustrated in Figure~\ref{figureparperp}b), where $\vec{v}_1$ and $\vec{v}_2$ are perpendicular.\footnote{Note that we must be clear what we mean by ``perpendicular,'' as recall that $\vec{v}_1$ is measured in mission control's rest frame, whilst $\vec{v}_2$ is measured in Alice's rest frame. It only makes sense to say two velocities are perpendicular if they are measured in the same reference frame. Hence when we say that $\vec{v}_1$ and $\vec{v}_2$ are perpendicular, we actually mean that, \textit{in Alice's frame of reference}, the velocity of mission control is $-\vec{v}_1$ and the velocity of Bob is $\vec{v}_2$ and these two velocities are perpendicular.} As there is no length contraction for perpendicular distances, but time dilation still occurs in moving from Alice's frame to mission control's, then the velocity $\vec{v}_2$ \textit{in mission control's reference frame} is just $\vec{v}_2/\gamma_1$. Therefore the velocity $\vec{v}_{21}$ of Bob as seen by mission control is just
\begin{equation}
 \vec{v}_{21}=\vec{v}_1+\frac{\vec{v}_2}{\gamma_1}=\vec{v}_1+\vec{v}_2\sqrt{1-v_1^2}\;.
 \label{perpv21}
\end{equation}
Similarly we find that
\begin{equation}
 \vec{v}_{12}=\vec{v}_2+\frac{\vec{v}_1}{\gamma_2}=\vec{v}_2+\vec{v}_1\sqrt{1-v_2^2}\;.
 \label{perpv12}
\end{equation}
{These formulae are extremely useful to introduce the concept of relativistically combining velocities. They are simple and almost trivial to derive, while still illustrating the fundamental concepts of relativistic velocity combination --- the non-intuitive addition laws, the Wigner rotation, and the Thomas precession --- as we shall now see.

\subsubsection{Wigner rotation}\label{perpwigsec}
Let us continue with the case of $\vec{v}_1$ and $\vec{v}_2$ being perpendicular, and hence with the relativistic combined velocities $\vec{v}_{21}$ and $\vec{v}_{12}$ as defined by \eqref{perpv21} and \eqref{perpv12} respectively. We note that
\begin{equation}
 \vec{v}_{21}\neq\vec{v}_{12},   \qquad\text{but}\qquad  ||\vec{v}_{12}||=||\vec{v}_{21}||=\sqrt{v_1^2+v_2^2-v_1^2v_2^2}\;.
 \label{perpmag}
\end{equation}
As $\vec{v}_{12}$ and $\vec{v}_{21}$ have the same magnitude, but a different direction, we expect some form of Wigner rotation, as previously discussed. One may naively guess that the Wigner rotation angle $\Omega$ may have something to do with the angle between $\vec{v}_{21}$ and $\vec{v}_{12}$, and in fact, as we will show in Section~\ref{angles}, it turns out that the angle between $\vec{v}_{21}$ and $\vec{v}_{12}$ is \textit{exactly} the Wigner rotation angle. We can easily calculate this angle by using the definition of the cross product and equations \eqref{perpv21}, \eqref{perpv12} and \eqref{perpmag}:
\begin{equation}
 \sin\Omega=\frac{||\vec{v}_{12}\times\vec{v}_{21}||}{||\vec{v}_{12}||\;||\vec{v}_{21}||}=\frac{v_1v_2\left(1-\frac{1}{\gamma_1\gamma_2}\right)}{v_1^2+v_2^2-v_1^2v_2^2}=\frac{v_1v_2\gamma_1\gamma_2}{1+\gamma_1\gamma_2}\;.
 \label{perpwig}
\end{equation}
Again, this is an extremely simple formula for the Wigner rotation angle $\Omega$, which is easily verifiable, using only the fundamental concepts of relativity. While \eqref{perpwig} only applies in the case of perpendicular velocities, it nonetheless introduces Wigner rotation, and is sufficient for considering the Thomas precession.
\subsubsection{Thomas precession}\label{perpthomsec}
Recall that the Thomas precession rate gives how fast Bob's frame is rotating with respect to mission control's. In our case, Alice sees mission control traveling at $-\vec{v}_1$ and Bob traveling at some infinitesimal velocity $\rmd{}\vec{v}_2$, as shown in Figure~\ref{figurethomassetup}b). If we assume that Bob is traveling in a circular orbit around Earth, then $\rmd{}\vec{v}_2$ is perpendicular to $\vec{v}_1$, and hence our formula for the Wigner rotation angle \eqref{perpwig} applies. As we let $\rmd{}\vec{v}_2\rightarrow{}0$, then $\gamma_{2}\rightarrow{}1$ and we find the infinitesimal Wigner rotation angle to first order in $\rmd{}{v}_2$ is (using the small angle approximation)
\begin{equation}
 \rmd{}\Omega=v_1\left(\frac{\gamma_1}{1+\gamma_1}\right)\rmd{}v_2\;.
 \label{perpthom}
\end{equation}
Hence the Thomas precession rate of Bob's frame as measured by mission control is
\begin{equation}
 \frac{\rmd{}\Omega}{\rmd{}t}=av_1\left(\frac{\gamma_1}{1+\gamma_1}\right)\;,
 \label{perpthomrate}
\end{equation}
where $a=\rmd{}v_2/\rmd{}t$ is the centripetal acceleration experienced by Bob. Hence we see that, at least for the specific case of circular motion, the formula describing the Thomas precession is simple, with a physically intuitive and elementary derivation.

\subsection{General velocities}\label{generalsec}
In general, the relativistic combination of velocities in arbitrary directions is nowhere near as simple as in the parallel and perpendicular cases previously discussed. However, we shall now present a derivation of a general formula for $\vec{v}_{21}$ which relies only on the elementary results of \eqref{parallel} and \eqref{perpv21}, and a simple time dilation argument. Let us consider the general situation of Figure~\ref{figureintro}, however we now decompose Bob's velocity as seen by Alice into its component $\vec{v}_{2\parallel{}1}$ parallel to $\vec{v}_1$ and its component $\vec{v}_{2\perp{}1}$ perpendicular to $\vec{v}_1$, as illustrated in Figure~\ref{figuredecomp}a). Let $S^{\,o}$ denote the rest frame of some contrived intermediate observer, whom Alice measures to have velocity $\vec{v}_{2\parallel{}1}$, as in Figure~\ref{figuredecomp}a).

\begin{figure}[!htb]
\begin{center}
\includegraphics[width=0.6\textwidth]{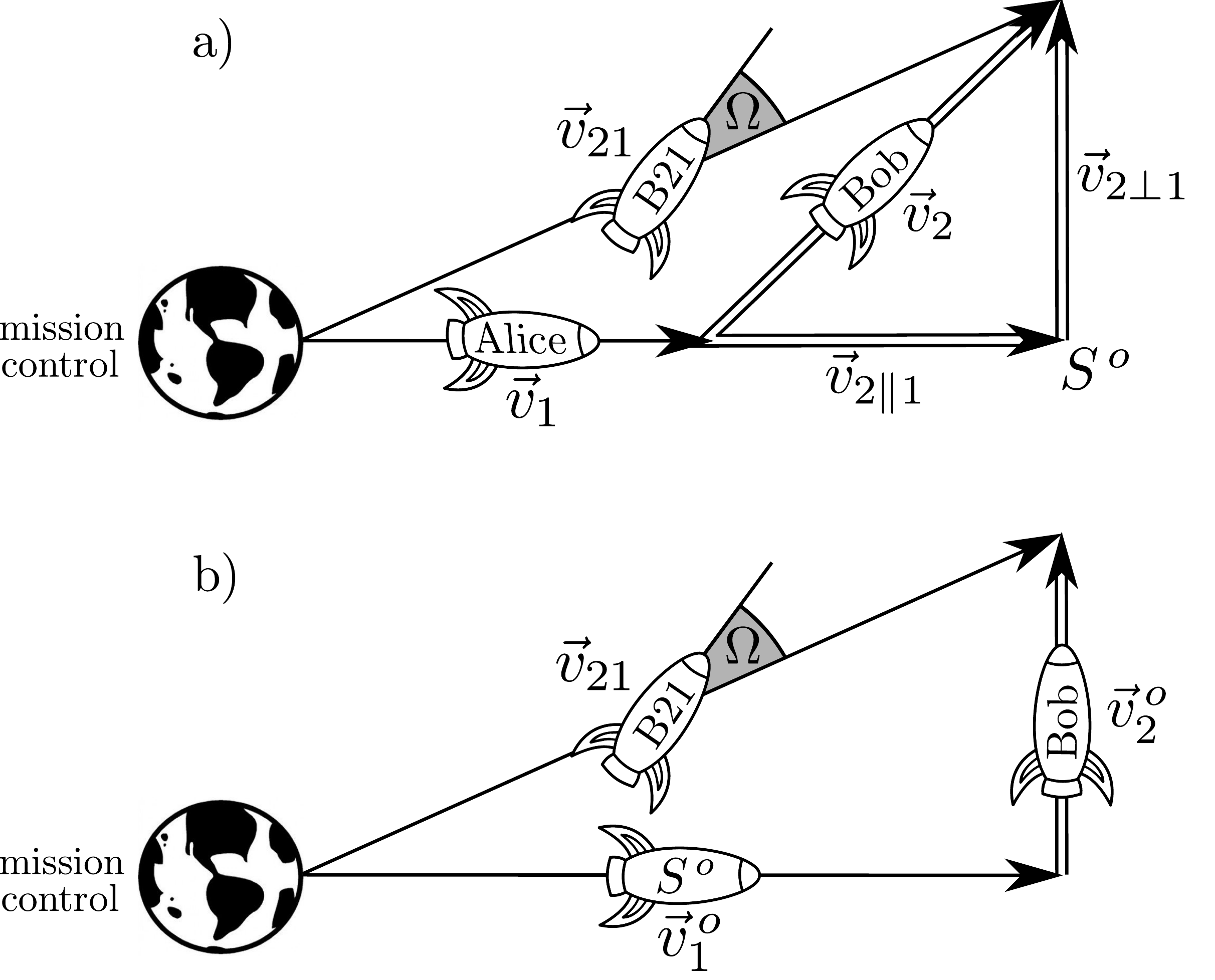}
\end{center}
\caption{
\textit{In subfigure a) we decompose the velocity $\vec{v}_2$ of Bob as measured by Alice into the components $\vec{v}_{2\parallel{}1}$ and $\vec{v}_{2\perp{}1}$, parallel and perpendicular to $\vec{v}_1$ respectively. The relativistically combined velocity $\vec{v}_{21}$ is the velocity of Bob as seen by mission control. In subfigure b), we see the $S^{\,o}$ frame, which is observed to have velocity $\vec{v}_1^{\,o}$ by mission control, which represents the relativistic combination of velocities $\vec{v}_{2\parallel{}1}$ and $\vec{v}_1$. In the $S^{\,o}$ frame Bob is measured to have velocity $\vec{v}_{2}^{\,o}$, perpendicular to $\vec{v}_1^{\,o}$.}
}
\label{figuredecomp}
\end{figure}

As $\vec{v}_{2\parallel{}1}$ and $\vec{v}_1$ are collinear, by \eqref{parallel}, the velocity of $S^{\,o}$ as measured by mission control is
\begin{equation}
 \vec{v}_1^{\,o}=\frac{\vec{v}_{2\parallel{}1}+\vec{v}_1}{1+\vec{v}_{2\parallel{}1}\cdot\vec{v}_1}=\frac{\vec{v}_{2\parallel{}1}+\vec{v}_1}{1+\vec{v}_{1}\cdot\vec{v}_2}\;,
 \label{parallel2}
\end{equation}
and there is no Wigner rotation of the $S^{\,o}$ frame relative to mission control. Therefore, we can think of a new situation, as illustrated in Figure~\ref{figuredecomp}b), where we have $S^{\,o}$ moving at velocity $\vec{v}_1^{\,o}$ relative to mission control, and Bob moving at some velocity $\vec{v}_{2}^{\,o}$ as measured in the $S^{\,o}$ frame. Using arguments similar to those in Section~\ref{elementary}, since $\vec{v}_{1}^{\,o}$ and $\vec{v}_{2}^{\,o}$ are perpendicular, then, due to time dilation
 \begin{equation}
 \vec{v}_{2}^{\,o}=\gamma_{2\parallel{}1}\; \vec{v}_{2\perp{}1}
 \qquad\text{where}\qquad
 \gamma_{2\parallel{}1}=\frac{1}{\sqrt{1-v_{2\parallel{}1}^2}}\;.
 \label{v2perp1'}
\end{equation}
 Thus, as mission control sees $S^{\,o}$ moving at velocity $\vec{v}_1^{\,o}$, and the observer $S^{\,o}$ sees Bob to be moving at the perpendicular velocity $\vec{v}_{2}^{\,o}=\gamma_{2\parallel{}1}\;\vec{v}_{2\perp{}1}$, we may apply formula \eqref{perpv21} for the relativistic combination of perpendicular velocities. Replacing $\vec{v}_1$ with $\vec{v}_1^{\,o}$ and $\vec{v}_2$ with $\vec{v}_{2}^{\,o}=\gamma_{2\parallel{}1}\;\vec{v}_{2\perp{}1}$ we see that the velocity of Bob with respect to mission control is given by
\begin{equation}
\vec{v}_{21}=\vec{v}_1^{\,o}+\frac{\gamma_{2\parallel{}1}}{\gamma_1^{\,o}}\; \vec{v}_{2\perp{}1}\;.
 \label{elemv21}
\end{equation}
Furthermore, from \eqref{parallel2} we see that 
\begin{equation}
 \gamma_1^{\,o}\equiv\frac{1}{\sqrt{1-(v_1^{\,o})^2}}=\gamma_{2\parallel{}1} \gamma_1(1+\vec{v}_{1}\cdot\vec{v}_2)\;,
\end{equation}
and hence \eqref{elemv21} becomes
\begin{equation}
 \vec{v}_{21}\;
 =\;\frac{\vec{v}_1+ \vec{v}_{2\parallel{}1}+\sqrt{1-v_1^2}\; \vec{v}_{2\perp{}1}}{1+\vec v_1\cdot{}\vec v_2}
  \;=\;\frac{\vec{v}_2+ \gamma_1 \vec{v}_1 +(\gamma_1 -1) (\vec{v}_1 \cdot \vec{v}_2) \vec{v}_1/v_1^2}{\gamma_1 (1 + \vec{v}_1 \cdot \vec{v}_2)}\;.
 \label{genv21}
\end{equation}
 Similarly, we find
 \begin{equation}
 \vec{v}_{12}\;
  =\;\frac{\vec{v}_2+ \vec{v}_{1\parallel{}2}+\sqrt{1-v_2^2}\; \vec{v}_{1\perp{}2}}{1+\vec v_1\cdot{}\vec v_2}
   \;=\;\frac{\vec{v}_1+ \gamma_2 \vec{v}_2 +(\gamma_2 -1) (\vec{v}_1 \cdot \vec{v}_2) \vec{v}_2/v_2^2}{\gamma_2 (1 + \vec{v}_1 \cdot \vec{v}_2)}\;.
 \label{genv12}
\end{equation} 
These are the most elementary formulae for the composition of general velocities that we have been able to uncover. Their derivation is simple and fundamental, with an easily attributable physical motivation.

\subsubsection{Wigner rotation}\label{genwigsec}
Whilst this subsection introduces no new concepts, the vector algebra becomes more tedious and may somewhat confuse the issue, so we consider this subsection to be more suitable for advanced students in a first course on relativity. We use a similar procedure as in Section~\ref{perpwigsec} to consider the Wigner rotation; to do so we must have $||\vec{v}_{21}||=||\vec{v}_{12}||$. We leave it to the reader to verify that indeed
\begin{equation}
||\vec{v}_{21}||=||\vec{v}_{12}||=\frac{\sqrt{||\vec{v}_1+\vec{v}_2||^2-||\vec{v}_1\times\vec{v}_2||^2}}{1+\vec{v}_1\cdot\vec{v}_2}\;,
\label{mag}
\end{equation}
and that this agrees with the parallel and perpendicular cases already discussed. Thus $\vec{v}_{21}$ and $\vec{v}_{12}$ have the same magnitude --- but by \eqref{genv21} and \eqref{genv12} they are not equal, and hence must point in different directions. As previously described for the perpendicular case in  Section~\ref{perpwigsec}, the Wigner rotation angle $\Omega$ is exactly the angle between $\vec{v}_{21}$ and $\vec{v}_{12}$ as measured by mission control. (We shall explicitly prove this in Section~\ref{angles}). To calculate $\Omega$, firstly rewrite \eqref{genv21} and \eqref{genv12} as
\begin{equation}
 \vec{v}_{21}=\frac{\vec{v}_1+ (1-\gamma_1^{-1}) \vec{v}_{2\parallel{}1} + \gamma_1^{-1}\vec{v}_{2}}{1+\vec{v}_{1}\cdot\vec{v}_2}\qquad\text{and}\qquad\vec{v}_{12}=\frac{\vec{v}_2+ (1-\gamma_{2}^{-1}) \vec{v}_{1\parallel{}2} + \gamma_{2}^{-1}\vec{v}_{1}}{1+\vec{v}_{1}\cdot\vec{v}_2}\;.
\label{v21v12diffb}
\end{equation}
The Wigner rotation angle $\Omega$ then follows from the cross--product of the vectors $\vec{v}_{21}$ and $\vec{v}_{12}$. Using \eqref{mag} and \eqref{v21v12diffb}, this results in
\begin{equation}
\sin(\Omega)=\frac{||\left(\vec{v}_2+ (1-\gamma_{2}^{-1}) \vec{v}_{1\parallel{}2} + \gamma_{2}^{-1}\vec{v}_{1}\right)\times \left(\vec{v}_1+ (1-\gamma_1^{-1}) \vec{v}_{2\parallel{}1} + \gamma_1^{-1}\vec{v}_{2}\right)||}{||\vec v_1+\vec v_{2}||^2  -  ||\vec v_1 \times \vec v_{2} ||^2}\;,
\label{sinomega2}
\end{equation}
which can be simplified to
\begin{equation}
\sin(\Omega) =v_1v_2\sin\theta{}\;\;
\frac{ \left[ 1- \gamma_1^{-1} \gamma_{2}^{-1} +   (\vec v_{1}\cdot \vec v_{2}) \left(\frac{1}{1+\gamma_1^{-1}}+\frac{1}{1+\gamma_{2}^{-1}} \right) + \frac{(\vec v_1\cdot \vec v_{2})^2}{(1+\gamma_1^{-1})(1+\gamma_{2}^{-1})}\right]}{||\vec v_1+\vec v_{2}||^2  -  ||\vec v_1 \times \vec v_{2} ||^2}\;,
\label{sinomega3}
\end{equation}
where $\theta$ is the angle between $\vec v_{1}$ and $\vec v_{2}$ as measured by Alice.\footnote{Note that $\vec{v}_1$ and $\vec{v}_2$ are in different frames, so it makes no sense to compare the angle between these velocities. What we really mean is that $\theta$ is the angle between $\vec v_{2}$ and the velocity, $-\vec{v}_{1}$, of mission control, as seen by Alice.} However
\begin{equation}
 \gamma_{12}\equiv\frac{1}{\sqrt{1-{v}_{12}^2}}=\gamma_1\gamma_2(1+\vec{v}_1\cdot{}\vec{v}_2)\;
 \label{gamma12}
\end{equation}
may be rearranged to give
\begin{equation}
 \cos\theta=\frac{\gamma_{12}-\gamma_1\gamma_{2}}{v_1v_2\gamma_1\gamma_{2}}\;.
\end{equation}
Hence, after a little massaging, \eqref{sinomega3} may then be simplified to
\begin{equation}
\sin\Omega = \frac{v_1v_{2}\gamma_1\gamma_{2} ( 1 + \gamma_1+ \gamma_{2} + \gamma_{12}  )}{(\gamma_1+1) (\gamma_2+1) (\gamma_{12}+1) }\; \sin\theta\;,
\label{stapp}
\end{equation}
which some may recognize as Stapp's elegant formula \cite{stapp1956}. Similarly, using the definition of the dot--product to find the Wigner rotation angle $\Omega$, one finds
\begin{equation}
 \cos\Omega=\frac{||\vec{v}_1+\vec{v}_{2}||^2-||\vec{v}_1\times\vec{v}_{2}||^2\left[\frac{1}{1+\gamma_1^{-1}}+\frac{1}{1+\gamma_{2}^{-1}}-\frac{\vec{v}_1\cdot{}\vec{v}_{2}}{\left(1+\gamma_1^{-1}\right)\left(1+\gamma_{2}^{-1}\right)}\right]}{||\vec v_1+\vec v_{2}||^2  -  ||\vec v_1 \times \vec v_{2} ||^2}\;,
\end{equation}
and eventually
\begin{equation}
{\cos\Omega+1}= \frac{(\gamma_{12}+ \gamma_1+\gamma_2+1)^2}{(\gamma_1+1)(\gamma_2+1)(\gamma_{12}+1) }\;.
\end{equation}
Indeed there are many explicit formulae for the Wigner rotation angle $\Omega$, a few of which are given in Section~\ref{summary}. Stapp's formula arguably remains the simplest and most useful. Note that while the derivation has been somewhat tedious in terms of algebra, the underlying physics is utterly elementary --- boiling down to the use of time dilation arguments combined with the usual composition of parallel velocities. 

\subsubsection{Thomas precession}\label{secthom}
We consider the same argument as given in Section~\ref{perpthomsec}, however we now do not make the simplifying assumption that $\vec{v}_1$ is perpendicular to $\rmd{}\vec{v}_2$ --- that is, Bob need not be in a circular orbit. It still remains true that the infinitesimal Wigner rotation in mission control's frame of reference is given by letting $\rmd{}\vec{v}_2\rightarrow{}0$, however we now use our general formula \eqref{stapp}. Doing so, then $\gamma_2\rightarrow{}1$, and from \eqref{gamma12}, we see that $\gamma_{12}\rightarrow\gamma_1$. Hence the infinitesimal Wigner rotation angle $\rmd{}\Omega$ is, to first degree in $\rmd{}\vec{v}_2$,
\begin{equation}
\rmd{}\Omega\approx\sin(\rmd{}\Omega) = v_1\rmd{}v_2\sin\theta\;\frac{\gamma_1}{1+\gamma_1}= ||\vec{v}_1\times{}\rmd{}\vec{v}_2||\;\frac{\gamma_1}{1+\gamma_1}\;.
\end{equation}
Therefore the Thomas precession rate in mission control's frame of reference is\footnote{There is a degree of confusion surrounding the precise definition of the Thomas precession. Our result agrees with that of \cite{malykin2006} and \cite{ritus2007}, and not with the more well known one of \cite{moller1952}, which gives the Thomas precession in mission control's frame as
\begin{equation}
\frac{\rmd{}\vec{\Omega}}{\rmd{}t}=\vec{v}_1\times{}\vec{a}\left(\frac{\gamma_1^2}{1+\gamma_1}\right)= \vec{v}_1\times{}\vec{a}\left(\frac{\gamma_1-1}{v_1^2}\right)\;.
\end{equation}
This however, as explained in \cite{malykin2006} and \cite{ritus2007}, is actually the Thomas precession rate as viewed from Alice's reference frame. The additional $\gamma_1$ factor is due to the time dilation between frames.} 
\begin{equation}
\frac{\rmd{}\vec{\Omega}}{\rmd{}t}=\vec{v}_1\times{}\vec{a}\left(\frac{\gamma_1}{1+\gamma_1}\right)\;,
\label{thomgen}
\end{equation}
where $\vec{a}=\rmd{}\vec{v}_2/\rmd{}t$ is the centripetal acceleration experienced by Bob. At this stage, it is clear that \eqref{thomgen} simplifies to the formula \eqref{perpthomrate} we obtained in the perpendicular case.

\section{Intermediate level --- boost matrix formulation}\label{intermediate}
We now consider the relativistic combination of velocities using the boost matrix formulation of special relativity. The results derived confirm those already found in Section~\ref{elementary}, however the use of boost matrices gives further conceptual insight --- notably that the Wigner rotation angle is the angle between $\vec{v}_{21}$ and $\vec{v}_{12}$.

\subsection{Composition of boosts}
Firstly, consider an arbitrary boost from a frame $S$ to another frame $S^{\,o}$ that is moving at a velocity $\vec{v}$ relative to $S$. Setting $c=1$, the boost matrix $B$ representing the transformation from the $S$ frame to the $S^{\,o}$ frame, such that $\vec{x}^{\,o}=B\vec{x}$, is
\begin{equation}
 B=\left[\begin{array}{c|c}
\vphantom{\Big{|}}\gamma & -\gamma\vec{v}^T \\ 
\hline\vphantom{\Big{|}}-\gamma\vec{v} & \mathbb{I}+(\gamma-1)\frac{{v}_i {v}_j}{{v}^2}\\
\end{array}\right]=\left[\begin{array}{c|c}\vphantom{\Big{|}}\gamma & -\gamma\vec{v}^T \\ \hline\vphantom{\Big{|}}-\gamma\vec{v} & P_v+\gamma{}Q_v\end{array}\right]\;,
\label{boost}
\end{equation}
where we have used the notation $P_v$ to represent the projection onto the plane perpendicular to $\vec{v}$ (explicitly, $[P_v]_{ij}=\delta_{ij}-v_iv_j/v^2$). Similarly $Q_v=\mathbb{I}-P_v$ gives the part parallel to $\vec{v}$. Now, any Lorentz transformation $L$ may be decomposed into a boost followed by a rotation:\footnote{Or a rotation followed by a boost --- we will use the form of \eqref{decomp} consistently throughout the paper.}
\begin{equation}
 L=RB
\label{decomp}
\end{equation}
for some rotation $R$ and some boost $B$. Furthermore, rotations take the form
\begin{equation}
 R = \left[ \begin{array}{c|c} 1 & 0 \\ \hline 0 & R_3 \end{array} \right] \;,
\label{rotation}
\end{equation}
where $R_3$ is some three-dimensional rotation matrix. Hence by \eqref{boost}, \eqref{decomp}, and \eqref{rotation}, any Lorentz transformation can be written in the form
\begin{equation}
L = \left[\begin{array}{c|c}  \vphantom{\Big{|}}  \gamma & - \gamma \vec v^T \\ \hline  \vphantom{\Big{|}}  -\gamma R_3 \vec v &  R_3 \{ P_v + \gamma\; Q_v  \} \end{array} \right]\;.
\label{decomp2}
\end{equation}
Thus we can calculate what the net Lorentz transformation $L_{21}$ is for the situation depicted in Figure~\ref{figureintro} by simply composing the two associated boosts --- that is, boosting first by $B_1$ and then $B_{2}$ --- so as to move from the $S$ frame to the $S^{\,o}$ frame. Hence
\begin{equation}
 L_{21}=B_{2}B_1\;.
\label{decomp0}
\end{equation}
Writing this out explicitly using \eqref{boost}, we see that
\begin{align}
L_{21}&=  
\left[\begin{array}{c|c}  \vphantom{\Big{|}}  \gamma_{2} & - \gamma_{2} \vec{v}_{2}^T \\ \hline  \vphantom{\Big{|}}  -\gamma_{2} \vec{v}_{2} &  P_2 + \gamma_{2}Q_2  \end{array} \right]
\;\;\;
\left[\begin{array}{c|c}  \vphantom{\Big{|}}  \gamma_1 & - \gamma_1 \vec{v}_1^T \\ \hline  \vphantom{\Big{|}}  -\gamma_1 \vec{v}_1 &  P_1 + \gamma_1Q_1  \end{array} \right]\\
&= \left[\begin{array}{c|c}  \vphantom{\Big{|}}  \gamma_{2}\gamma_1(1+\vec{v}_{2}\cdot\vec{v}_1)  & 
 - \gamma_{2}\gamma_1\vec{v}_1^T -\gamma_{2}\vec{v}_{2}^T[P_1 + \gamma_1Q_1  ]
 \\ \hline  
 \vphantom{\Big{|}}  -\gamma_{2}\gamma_1\vec{v}_{2} -\gamma_1[P_2 + \gamma_{2}Q_2 ] \vec{v}_1
 &  
[P_2 + \gamma_{2}Q_2 ] [ P_1 + \gamma_1Q_1] +\gamma_1\gamma_{2} \vec{v}_{2} \vec{v}_1^T \end{array} \right]\,.
\label{comp}
\end{align}
Thus if we wish to decompose this Lorentz transformation into the form $L_{21}=RB_{21}$, and we let $\vec{v}_{21}$ denote the velocity corresponding to the boost $B_{21}$, we can equate \eqref{decomp2} and \eqref{comp}, which gives
\begin{equation}
\left[\begin{array}{c|c}  \vphantom{\Big{|}}  \gamma_{21} & - \gamma_{21} \vec v_{21}^T \\ \hline  \vphantom{\Big{|}}  -\gamma_{21} R_3 \vec{v}_{21} &  R_3 [ P_{21} + \gamma\; Q_{21}  ] \end{array} \right]\qquad\qquad\qquad\qquad\qquad\qquad\qquad\qquad\qquad\qquad\qquad\nonumber
\end{equation}
\begin{equation}
 \qquad= \left[\begin{array}{c|c}  \vphantom{\Big{|}}  \gamma_{2}\gamma_1(1+\vec{v}_{2}\cdot\vec{v}_1)  & 
 - \gamma_{2}\gamma_1\vec{v}_1^T -\gamma_{2}\vec{v}_{2}^T[P_1 + \gamma_1Q_1  ]
 \\ \hline  
 \vphantom{\Big{|}}  -\gamma_{2}\gamma_1\vec{v}_{2} -\gamma_1[P_2 + \gamma_{2}Q_2 ] \vec{v}_1
 &  
[P_2 + \gamma_{2}Q_2 ] [ P_1 + \gamma_1Q_1] +\gamma_1\gamma_{2} \vec{v}_{2} \vec{v}_1^T \end{array} \right]\,.
\label{bigcomp}
\end{equation}
Comparing the $00$ terms, we see that
\begin{equation}
\gamma_{21}=\gamma_{2}\gamma_1(1+\vec{v}_{2}\cdot\vec{v}_1)\;,
\label{gamma21}
\end{equation}
as we found previously in \eqref{gamma12} (note that $\gamma_{21}=\gamma_{12}$). Using this result in comparing the $0j$ terms of \eqref{bigcomp}, we see that
\begin{equation}
\vec{v}_{21}=\frac{\vec{v}_1 + \gamma_1^{-1}P_1\vec{v}_{2} + Q_1\vec{v}_{2}}{1+\vec{v}_{2}\cdot\vec{v}_1}\;.
\end{equation}
This can be written alternatively as
\begin{equation}
\vec{v}_{21}=\frac{\vec{v}_1+ \vec{v}_{2\parallel{}1} + \sqrt{1-v_1^2}\;\vec{v}_{2\perp{}1}}{1+\vec{v}_{2}\cdot\vec{v}_1}\;,
\label{v21}
\end{equation}
which is what was derived in Section~\ref{generalsec}. Furthermore, as follows from \eqref{decomp} and \eqref{decomp0}, we can define a pure rotation matrix $R$ and a pure boost matrix $B_{21}$ such that
\begin{equation}
 B_{2}B_1=RB_{21}\;.
 \label{boost00}
\end{equation}
However for the same rotation matrix $R$, we have
\begin{equation}
 B_{1}B_{2}=(B_{2}B_1)^T=(RB_{21})^T=B_{21}^TR^T=B_{21}R^{-1}=R^{-1}(RB_{21}R^{-1})\;.
\label{boostcomp}
\end{equation}
Since $(RB_{21}R^{-1})^T=RB_{21}R^{-1}$, we see that $RB_{21}R^{-1}$ is a pure boost --- so we define $B_{12}\equiv{}RB_{21}R^{-1}$, such that
\begin{equation}
 B_{1}B_{2}=R^{-1}B_{12}\;.
\label{boost1}
\end{equation}
The results \eqref{boost00} and \eqref{boost1} verify the interpretation given in Figure~\ref{figurev12v21}, that whilst mission control may measure Bob to be moving with velocity $\vec{v}_{21}$, his frame of reference will be rotated by $\Omega$, and similarly for $\vec{v}_{12}$ (except the rotation will be by $-\Omega$). Furthermore, from \eqref{boost1} it follows that
\begin{equation}
 R=B_{12}B_{2}^{-1}B_1^{-1}\;,
\label{boost2}
\end{equation}
and hence we have explicitly calculated the rotation matrix $R$. We can ``simplify'' this further however, by using \eqref{boost1} and noting that
\begin{equation}
B_{2}B_1B_1B_{2}=B_{12}^2\ .
\end{equation}
Thus by \eqref{boost2}
\begin{equation}
 R=\sqrt{B_{2}B_1B_1B_{2}}B_{2}^{-1}B_1^{-1}\;.
\end{equation}
\\[-0.5cm]
We can now use the property that the angle $\Omega$ rotated by in the rotation $R$ is related to the trace of the rotation matrix via $\text{tr}(R)=2(1+\cos(\Omega))$, and hence the Wigner rotation angle $\Omega$ is
\begin{equation}
 \cos\Omega + 1=\frac{1}{2}\text{tr}\left(\sqrt{B_{2}B_1B_1B_{2}}B_{2}^{-1}B_1^{-1}\right)\;.
\end{equation}
%
%
\subsection{Connecting the angles}\label{angles}

In deriving the formulae in Section~\ref{perpwigsec} and Section~\ref{genwigsec} for the Wigner rotation angle, we assumed that it was the angle between $\vec{v}_{21}$ and $\vec{v}_{12}$. We can now prove this using our boost matrix formulation.\\
\\
By \eqref{decomp} and \eqref{rotation}, the Wigner rotation angle $\Omega$ is just the angle involved in the rotation $R$, or equivalently, the three-dimensional rotation $R_3$. However, consider again \eqref{genv12} and \eqref{gamma21}, which show that
\begin{equation}
\gamma_{21}\vec{v}_{12}=\gamma_{2}\gamma_1\vec{v}_{2} +\gamma_1[P_2 + \gamma_{2}Q_2 ] \vec{v}_1\;.
\label{yrv}
\end{equation}
By equating the $i0$ entries of \eqref{bigcomp}, we see that $\gamma_{21}R_3\vec{v}_{21}$ is also equal to the right-hand-side of \eqref{yrv}, and hence
\begin{equation}
 R_3 \vec{v}_{21}=\vec{v}_{12}\;.
\end{equation}
Therefore, as previously claimed and now proved, we can find the Wigner rotation angle $\Omega$ by simply calculating the angle between $\vec{v}_{12}$ and $\vec{v}_{21}$.
\subsection{Inverting the transformations}
As a final consideration, what is the velocity of mission control as seen by Bob? If mission control sees Bob moving with velocity $\vec{v}_{21}$, does Bob see mission control moving with velocity $-\vec{v}_{21}$ as would be expected in Galilean relativity? If Bob's frame is considered as the observer frame, then he will see Alice moving with velocity $-\vec{v}_2$ and Alice will see mission control moving with velocity $-\vec{v}_1$. Hence the velocity of mission control as observed by Bob is given by the composition of the two velocities $-\vec{v}_2$ and then $-\vec{v}_1$, or in boost matrix form
\begin{equation}
 B_{-1}B_{-2}=B_1^{-1}B_2^{-1}=(B_2B_1)^{-1}=(RB_{21})^{-1}=B_{21}^{-1}R^{-1}\;,
\label{boostcomp2}
\end{equation}
where we have used that $B_{-v}=B_v^{-1}$, i.e. the inverse of a boost in a direction $\vec{v}$ is just a boost in the direction $-\vec{v}$. However from our definition of $B_{12}\equiv{}RB_{21}R^{-1}$, we see that $B_{21}=R^{-1}B_{12}R$ and hence $B_{21}^{-1}=R^{-1}B_{12}^{-1}R$, so \eqref{boostcomp2} implies that
\begin{equation}
 B_{-1}B_{-2}=R^{-1}B_{12}^{-1}\;.
\label{boostcomp3}
\end{equation}
Thus the transformation from Bob's frame to mission control's frame is given by
\begin{equation}
B_{-1}B_{-2}=B_{-21}R^{-1}=R^{-1}B_{-12}\;,
\end{equation}
where $B_{-12}$ and $B_{-21}$ correspond to boosts in the $-\vec{v}_{12}$ and $-\vec{v}_{21}$ directions respectively. However which one, $B_{-21}R^{-1}$ or $R^{-1}B_{-12}$, should we consider to determine the velocity of mission control as observed by Bob? The transformation $B_{-21}R^{-1}$ implies \textit{first} rotating Bob's frame, and \textit{then} boosting along $-\vec{v}_{21}$ to end up in mission control's frame. Hence \textit{in Bob's rotated frame} he will see mission control traveling at velocity $-\vec{v}_{21}$. However the transformation $R^{-1}B_{-12}$ implies first boosting from Bob's frame by $-\vec{v}_{12}$, and \textit{then} rotating. Therefore in Bob's \textit{original frame}, he will see mission control traveling with velocity $-\vec{v}_{12}$ (but pointing in a direction rotated by $-\Omega$, due to Wigner rotation). Thus while mission control sees Bob moving with velocity $\vec{v}_{21}$ in their frame of reference, Bob sees mission control moving with velocity $-\vec{v}_{12}$ in his.

\section{Advanced level --- spinor formulation}\label{advanced}

\def\sech{{\mathrm{sech}}}
\def\tr{{\mathrm{tr}}}
\def\s{\boldsymbol{{\sigma}}}
\def\x{\mathbf{{x}}}
\def\n{\mathbf{{n}}}
\def\a{\mathbf{{a}}}
\def\b{\mathbf{{b}}}
\def\v{\mathbf{{v}}}
\def\u{\mathbf{{u}}}
\def\X{\mathbf{{X}}}
\def\B{\mathbf{{B}}}
\def\R{\mathbf{{R}}}
In this section we use the spinor formulation of special relativity to consider the relativistic combination of velocities and the Wigner rotation angle. There are in fact two methods of finding the results we require. The first is straightforward but tedious, so we only present the initial formulation, and the final results. The second version is less apparent, but much quicker, and we give the full derivation.
\subsection{Explicit approach}

Let $\sigma = (\sigma_x, \sigma_y, \sigma_z)$ be a 3-vector of Pauli sigma matrices, and let
\begin{equation}
\X = ct I + \x \cdot \s 
\end{equation}
be a representation of a 4-vector $\X=(ct,\x)$ in terms of a Hermitian $2\times2$ matrix. Then boosts are represented by
\begin{equation}
\X \to \B \X \B;   \qquad   \B = \cosh(\xi/2) +  \sinh(\xi/2) \; \n \cdot \s\;;
\end{equation}
and rotations by
\begin{equation}
\X \to \R \X \R^{-1};  \qquad   \R = \cos(\theta/2) +  i \sinh(\theta/2) \; \n \cdot \s\;.
\end{equation}
Where $\xi$ is the rapidity parameter defined by $v=\tanh \xi$. The Wigner rotation is now encoded in the fact that
\begin{equation}
\B_2 \B_1 = \R_{21} \B_{21}\;.
\end{equation}
One can write this out explicitly. Defining $\vec \Omega = \Omega \; \vec n_\Omega$, upon equating coefficients, we find the following four simultaneous, independent equations:
\begin{equation}
 \n_\Omega \cdot \n_{12} = 0\;,
\end{equation}
\begin{equation}
 \cos\frac{\Omega}{2} \cosh \frac{\xi_{12}}{2}  = \cosh \frac{\xi_{2}}{2}  \cosh \frac{\xi_{1}}{2}  + \sinh \frac{\xi_{2}}{2}  \sinh \frac{\xi_{1}}{2}  \; \n_2 \cdot \n_1\;,
\end{equation}
\begin{align}
 \cos \frac{\Omega}{2}   \sinh \frac{\xi_{12}}{2}  \; \n_{12} - \sin \frac{\Theta}{2}   \sinh \frac{\xi_{12}}{2}    \n_\Omega \times\n_{12} & \nonumber \\
= \cosh \frac{\xi_{2}}{2}   \sinh \frac{\xi_{1}}{2}  \; \n_1 + \cosh \frac{\xi_{1}}{2} &  \sinh \frac{\xi_{2}}{2}  \; \n_2\;,
\end{align}
\begin{equation}
 \cosh \frac{\xi_{12}}{2}   \sin \frac{\Omega}{2}  \; \n_\Omega  =  \sinh \frac{\xi_{1}}{2}   \sinh \frac{\xi_{2}}{2}   \;  \n_1\times\n_2 \;.
\end{equation}
One can then test one's algebraic skill and fortitude, to eventually arrive at the already proven results for the Wigner rotation:
 \begin{equation}
{\cos\Omega+1}= \frac{(\gamma_{12}+ \gamma_1+\gamma_2+1)^2}{(\gamma_1+1)(\gamma_2+1)(\gamma_{12}+1) }\;,
\end{equation}
and the familiar formula of Stapp~\cite{stapp1956}
\begin{equation}
\sin\Omega = \frac{v_1\gamma_1v_2\gamma_2 ( 1 + \gamma_1+ \gamma_2 + \gamma_{12}  )} {(\gamma_1+1) (\gamma_2+1) (\gamma_{12}+1) }\, \sin\theta\;.
\end{equation}
We congratulate those who verify this procedure! All the physics is already encoded in the equations above --- the only difficulty lies in the tedious nature of the algebra.

\subsection{A more efficient approach}
\def\tr{\hbox{tr}}
Whilst this derivation is significantly shorter step-wise, it involves some not entirely obvious leaps of understanding that we leave for the reader to verify. To begin with
\begin{equation}
\B_{12}^2 =   \B_1\B_2\B_2\B_1\;,
\end{equation}
and hence
\begin{equation}
\gamma_{12} = \frac{1}{2} \tr(\B_{12}^2) = \frac{1}{2}\tr(\B_1\B_2\B_2\B_1) = \frac{1}{2} \tr( \B_1^2 \B_2^2) = \gamma_1\gamma_2 (1+\v_1\cdot\v_2)\;.
\end{equation}
This then leads to
\begin{equation}
\cos\theta = \frac{\gamma_{12}-\gamma_1\gamma_2}{\sqrt{(\gamma_1^2-1)(\gamma_2^2-1)}}\;.
\end{equation}
Using these results, and the fact that $\tr(\R_{12} \B_{12}) = \tr(\B_1 \B_2)$, we find
\begin{equation}
\sqrt{1+\cos\Omega} \sqrt{\gamma_{12}+1}= \sqrt{(\gamma_1+1)(\gamma_2+1)} + \sqrt{(\gamma_1-1)(\gamma_2-1)} \cos\theta\;.
\end{equation}
Thus
\begin{equation}
\cos\Omega+1 = \frac{ (1+\gamma_1+\gamma_2+ \gamma_{12})^2}{(\gamma_1+1)(\gamma_2+1)(\gamma_{12}+1)}
\end{equation}
as required.
\section{Summary of useful formulae}\label{summary}

\paragraph{General:} The relativistic combination of general velocities $\vec{v}_1$ and $\vec{v}_2$:
\begin{equation}
 \vec{v}_{21}\;
 =\;\frac{\vec{v}_1+ \vec{v}_{2\parallel{}1}+\sqrt{1-v_1^2}\; \vec{v}_{2\perp{}1}}{1+\vec v_1\cdot{}\vec v_2}
  \;=\;\frac{\vec{v}_2+ \gamma_1 \vec{v}_1 +(\gamma_1 -1) (\vec{v}_1 \cdot \vec{v}_2) \vec{v}_1/v_1^2}{\gamma_1 (1 + \vec{v}_1 \cdot \vec{v}_2)}\;,
\end{equation}
 \begin{equation}
 \vec{v}_{12}\;
  =\;\frac{\vec{v}_2+ \vec{v}_{1\parallel{}2}+\sqrt{1-v_2^2}\; \vec{v}_{1\perp{}2}}{1+\vec v_1\cdot{}\vec v_2}
   \;=\;\frac{\vec{v}_1+ \gamma_2 \vec{v}_2 +(\gamma_2 -1) (\vec{v}_1 \cdot \vec{v}_2) \vec{v}_2/v_2^2}{\gamma_2 (1 + \vec{v}_1 \cdot \vec{v}_2)}\;,
\end{equation} 
\begin{equation}
||\vec{v}_{21}||=||\vec{v}_{12}||=\frac{\sqrt{||\vec{v}_1+\vec{v}_2||^2-||\vec{v}_1\times\vec{v}_2||^2}}{1+\vec{v}_1\cdot\vec{v}_2}\;,
\label{mag2}
\end{equation}
\begin{equation}
 \gamma_{12}=\gamma_1\gamma_2(1+\vec{v}_1\cdot{}\vec{v}_2)\;.
\end{equation}
\\
The Wigner rotation angle $\Omega$:
\begin{equation}
\sin\Omega = \frac{v_1\gamma_1v_2\gamma_2 ( 1 + \gamma_1+ \gamma_2 + \gamma_{12}  )} {(\gamma_1+1) (\gamma_2+1) (\gamma_{12}+1) } \sin\theta\;,
\end{equation}
\begin{equation}
{\cos\Omega+1}= \frac{(\gamma_{12}+ \gamma_1+\gamma_2+1)^2}{(\gamma_1+1)(\gamma_2+1)(\gamma_{12}+1) }\;.
\end{equation}
\\
The Thomas precession as seen in mission control's reference frame:
\begin{equation}
\frac{\rmd{}\vec{\Omega}}{\rmd{}t}=\vec{v}_1\times{}\vec{a}\left(\frac{\gamma_1}{1+\gamma_1}\right)\;.
\end{equation}
\\
The Thomas precession as seen in Alice's reference frame:
\begin{equation}
\frac{\rmd{}\vec{\Omega}}{\rmd{}t}=\vec{v}_1\times{}\vec{a}\left(\frac{\gamma_1^2}{1+\gamma_1}\right)\;.
\end{equation}

\paragraph{Perpendicular:} The relativistic combination of perpendicular velocities $\vec{v}_1$ and $\vec{v}_2$ is particularly elegant:
\begin{equation}
 \vec{v}_{21}=\vec{v}_1+\sqrt{1-v_1^2}\;\vec{v}_{2}\;,
\end{equation}
\begin{equation}
 \vec{v}_{12}=\vec{v}_2 + \sqrt{1-v_2^2}\;\vec{v}_{1}\;,
\end{equation}
\begin{equation}
||\vec{v}_{21}||=||\vec{v}_{12}||=\sqrt{v_1^2+v_2^2-v_1^2 v_2^2}\;,
\label{mag2c}
\end{equation}
\begin{equation}
 \gamma_{12}=\gamma_1\gamma_2\;.
\end{equation}
\\
The Wigner rotation angle $\Omega$:
\begin{equation}
\sin\Omega = \frac{v_1\gamma_1v_2\gamma_2} {\gamma_1\gamma_2+1} \;,
\end{equation}
\begin{equation}
{\cos\Omega+1}= \frac{(\gamma_1+1)(\gamma_2+1)}{\gamma_{1}\gamma_2+1 }\;.
\end{equation}

\section{Further reading}\label{furtherreading}
For those students interested in more details regarding the relativistic combination of velocities from a reasonably  elementary viewpoint, the explicit boost composition approach taken in reference \cite{ferraro1999} may prove useful. There are then many other standard textbook approaches, such as can be found in \cite{moller1952} or \cite{jackson1998}. Some of the finer details about the relativistic combination of velocities, especially the relationships between the different frames, can be found in references \cite{fisher1972} and \cite{ritus2007}.

However the issue that receives the most attention in the literature is the Thomas precession, (and to a lesser extent the associated Wigner rotation) --- partly due to the confusion surrounding it. We feel that readers further interested in the Thomas precession (and relativistic velocity combination in general), will benefit greatly from reference \cite{malykin2006}, which gives both a review of the literature (where the reader can find many higher-level approaches outlined), a select few of the possible derivations of the Thomas precession formula and some physical interpretations, and it also clarifies some of the misconceptions surrounding the Thomas precession. One such of these is what actually is the correct formulation, as alluded to in Section~\ref{secthom}. Reference \cite{ritus2007} provides further discussion on this point.



\begin{thebibliography}{69}

\bibitem{moller1952}
C.~M{\o}ller.
{\em The Theory of Relativity}.
(Oxford University Press, London, 1952).

\bibitem{jackson1998}
J.~D. Jackson.
{\em Classical Electrodynamics, 3rd Ed.}
(Wiley, New York, 1998).


\bibitem{stapp1956}
H.~P. Stapp.
``Relativistic theory of polarization phenomena.''
{\em Physical Review}, {\bf 103} (1956) 425--434.

\bibitem{fisher1972}
G.~P. Fisher.
``Thomas precession.''
{\em American Journal of Physics}, {\bf 40} (1972) 1772.

\bibitem{ferraro1999}
M.~Ferraro, R.~Thibeault.
``Generic composition of boosts: an elementary derivation of the Wigner
  rotation.''
{\em European Journal of Physics}, {\bf 20} (1999) 143.

\bibitem{malykin2006}
G.~B. Malykin.
``Thomas precession: correct and incorrect solutions.''
{\em Physics--Uspekhi}, {\bf 49} (2006) 837--853.

\bibitem{ritus2007}
V.~I. Ritus.
``On the difference between Wigner's and M\o{}ller's approaches to the
  description of Thomas precession.''
{\em Physics--Uspekhi}, {\bf 50} (2007) 95--101.




\end{thebibliography}
\end{document}